\documentclass[english,aps,prl,twocolumn,preprintnumbers,superscriptaddress,nofootinbib]{revtex4-1}
\usepackage[T1]{fontenc}
\usepackage[latin9]{inputenc}
\usepackage{amsmath}
\usepackage{graphicx}

\makeatletter
\usepackage[colorlinks=true,citecolor=blue,linkcolor=blue]{hyperref}
\usepackage[normalem]{ulem}
\usepackage{amsmath,amssymb}
\usepackage{mathtools}
\usepackage{verbatim}
\usepackage{titlesec}               
\usepackage{epsfig}
\usepackage{graphicx}               
\usepackage{url}
\usepackage{color}
\usepackage{multirow}
\usepackage{placeins}
\usepackage[dvipsnames]{xcolor}
\usepackage{epstopdf}
\usepackage{enumitem}
\usepackage[capitalize]{cleveref}
\usepackage{lipsum}
\usepackage{gensymb}

\allowdisplaybreaks

\makeatletter
\renewcommand{\thesection}{\arabic{section}}

\renewcommand{\p@subsection}{}
\renewcommand{\p@subsubsection}{}
\makeatother

\titleformat*{\section}{\centering\bfseries\uppercase}
\titlelabel{\thetitle\quad}
\titleformat*{\paragraph}{\bfseries}
\titlespacing*{\paragraph}{0pt}{3.25ex plus 1ex minus .2ex}{1em}

\makeatletter
\def\l@subsubsection#1#2{}
\makeatother

\makeatother

\usepackage{babel}
\begin{document}
\title{Beyond Tree Level with Solar Neutrinos:\\
Towards Measuring the Flavor Composition and CP Violation}

\author{Vedran Brdar}
\email{vedran.brdar@cern.ch}
\affiliation{Theoretical Physics Department, CERN,
             1 Esplanade des Particules, 1211 Geneva 23, Switzerland}

\author{Xun-Jie Xu}
\email{xuxj@ihep.ac.cn}
\affiliation{Institute of High Energy Physics, Chinese Academy of Sciences, Beijing 100049, China}

\preprint{CERN-TH-2023-097}

\begin{abstract}
\noindent
After being produced as electron neutrinos ($\nu_e$), solar neutrinos partially change their flavor to $\nu_{\mu}$ and $\nu_{\tau}$ en route to Earth.  Although the flavor ratio of the $\nu_e$ flux to the total flux has been well measured, the $\nu_{\mu}:\nu_{\tau}$ composition has not yet been experimentally probed. In this work we investigate the potential of the next-generation experiments for measuring the $\nu_{\mu}:\nu_{\tau}$ flavor ratio by utilizing flavor-dependent radiative corrections in the cross sections for $\nu_{\mu}$ and $\nu_{\tau}$ scattering. Since the transition probabilities of $\nu_e$ to $\nu_\mu$ and $\nu_\tau$ depend on the leptonic CP phase, we also investigate the sensitivity to the CP phase and show that a statistical significance of $\sim1 \sigma$ could be reached through precision measurements of solar neutrino spectra. 
\end{abstract}

\maketitle

\textbf{Introduction.}
The first observation of solar neutrinos at the Homestake experiment~\cite{Davis:1968cp} was not consistent with the theoretical predictions from Bahcall~\cite{Bahcall:1968hc} and this turned out to be the first experimental hint for neutrino oscillation~\cite{Pontecorvo:1967fh,Gribov:1968kq,Wolfenstein:1977ue,Mikheev:1986gs,Mikheev:1986wj}. This phenomenon implies that neutrinos produced in the Sun change flavor en route to Earth, which has by now been confirmed with a number of experiments including SNO \cite{SNO:2001kpb,SNO:2002tuh}, Super-Kamiokande \cite{Super-Kamiokande:2001ljr,Super-Kamiokande:2002ujc}, and Borexino \cite{Borexino:2008dzn,BOREXINO:2020aww}. For recent reviews on solar neutrino physics, see \cite{Maltoni:2015kca,Xu:2022wcq}.

The modern-day solar neutrino observations have established that only about a third of neutrinos produced in the Sun arrive to Earth as electron neutrinos ($\nu_{e}$),
while the remaining fraction is composed of muon ($\nu_{\mu}$) and
tau neutrinos ($\nu_{\tau}$). The ratio between $\nu_{\mu}$ and $\nu_{\tau}$ fluxes
is theoretically known but it has never been measured. 
A full measurement of the flavor composition ($\nu_{e}:\nu_{\mu}:\nu_{\tau}$)
would be valuable, as it would not only allow us to gain a better
understanding of our nearest star, but would also make it possible to probe
the transition probabilities $P_{\nu_e \to \nu_{\mu}}$ and $P_{\nu_e \to \nu_{\tau}}$ which depend on the level of the CP violation in the lepton sector (parameterized
by the phase $\delta_{\rm CP}$). 
Given such dependence,  solar neutrinos could serve as a complementary
probe of CP violation to the near-future acceleration-based neutrino
program led by DUNE \cite{DUNE:2015lol} and Hyper-Kamiokande (HK) \cite{Abe:2011ts}.

\begin{figure}
	\centering \includegraphics[width=9.3cm]{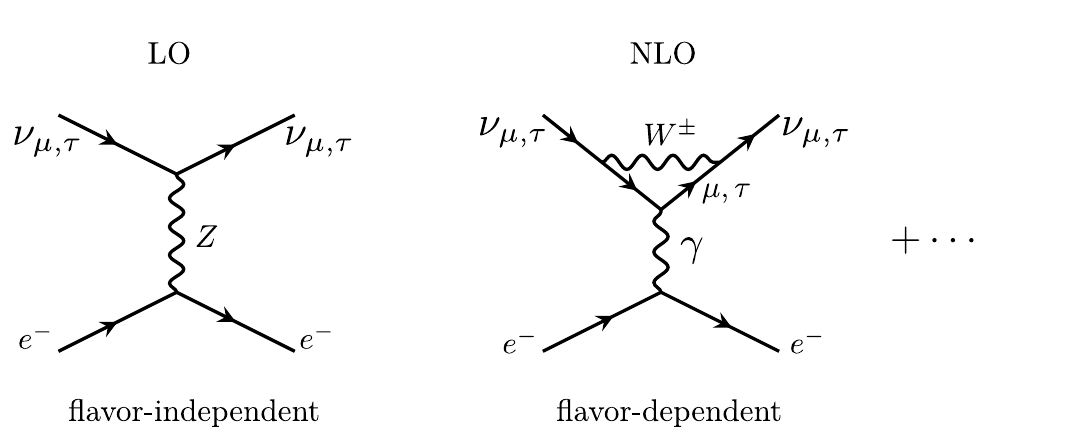} \caption{Feynman diagrams for $\nu_{\mu,\tau}$ scattering on electrons. The
		diagram on the left shows the leading order contribution whereas the
		one on the right illustrates the flavor-dependent NLO contribution.}
	\label{fig:feynman} 
\end{figure}

In this letter we propose a viable method to differentiate
between solar $\nu_{\mu}$ and $\nu_{\tau}$. Since solar neutrino
energies do not exceed $\sim 20$ MeV, $\nu_{\mu}$ and $\nu_{\tau}$
can only be detected via elastic scattering and, at the leading order, cross sections for this process are identical for both $\nu_{\mu}$ and $\nu_{\tau}$.   
At the next-to-leading order (NLO), however, differences arise from flavor-dependent radiative corrections ~\cite{Sehgal:1985iu,Degrassi:1989ip,Tomalak:2020zfh}; see e.g.  \cref{fig:feynman}.

In this work we will be focused on solar neutrinos and the phenomenological consequences that radiative corrections in the cross section for neutrino-electron elastic scattering (eES) \cite{Sarantakos:1982bp,Marciano:2003eq} can induce. One of the advantages for considering $\nu_{\mu,\tau}+e^{-}\to\nu_{\mu,\tau}+e^{-}$
is the rather small theoretical uncertainty in the cross section which is at sub-percent level \cite{Tomalak:2019ibg,Hill:2019xqk}.
With sufficiently high statistics, the difference between $\nu_{\mu}$
and $\nu_{\tau}$ cross sections could manifest itself in the data, namely
in the total number of eES events. We investigate
that by considering large next-generation neutrino detectors such
as HK \cite{Abe:2011ts}, DUNE \cite{DUNE:2015lol},
JUNO \cite{JUNO:2015zny} and THEIA \cite{Theia:2019non}.

\begin{figure}
	\centering 
	
	\includegraphics[width=7cm]{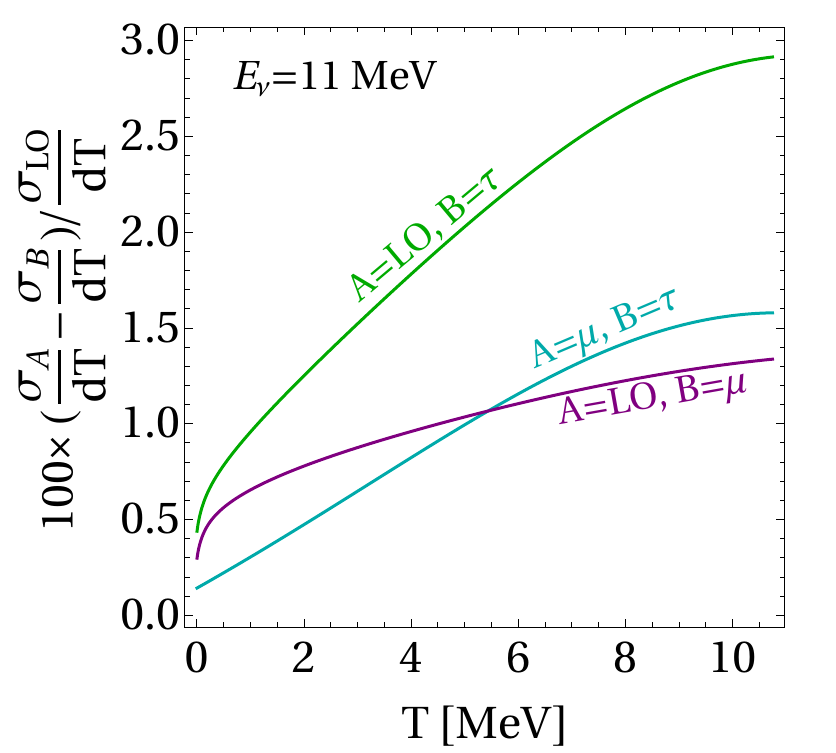} \caption{Comparison between the differential cross sections, normalized to
		the leading order one (see \cref{eq:tree}). In particular, we would
		like to draw reader's attention to the cyan line that demonstrates $\mathcal{O}$(1\%)
		difference between $\nu_{\mu}$ and $\nu_{\tau}$ scattering cross
		sections. This is the effect our phenomenological study is based on.}
	\label{fig:xsec} 
\end{figure}

\textbf{Radiative corrections for eES cross section.}  As already introduced, in this work we will mainly consider
eES, the experimental signature of which is the detection of electron
recoil kinetic energy, denoted as $T$. The differential cross section
without the inclusion of any radiative corrections reads \cite{Marciano:2003eq} 
\begin{align}
\frac{d\sigma}{dT} & =\frac{2G_{F}^{2}m_{e}}{\pi}\bigg[\left(s_{W}^{2}\pm\frac{1}{2}\right)^{2}+s_{W}^{4}\left(1-\frac{T}{E_{\nu}}\right)^{2}-\nonumber \\
 & \left(s_{W}^{2}\pm\frac{1}{2}\right)s_{W}^{2}\frac{m_{e}T}{E_{\nu}^{2}}\bigg]\,,\label{eq:tree}
\end{align}
where the $+$ $(-)$ sign applies for $\nu_{e}$ ($\nu_{\mu}$ or
$\nu_{\tau}$) scattering on electron. In \cref{eq:tree}, $G_{F}$
is the Fermi constant, $s_{W}$ stands for the sine of the weak mixing
angle and $m_{e}$ is the electron mass.

The difference between $\nu_{\mu}$ and $\nu_{\tau}$ scattering cross
sections comes about at NLO \cite{Sarantakos:1982bp,Marciano:2003eq,Sirlin:2012mh}.
We are in particular interested in flavor-dependent corrections, 
which arise from higher-order diagrams involving $\mu$ and $\tau$ leptons in the loops, such as the right diagram in \cref{fig:feynman}.
This effect can be accounted for via redefinition $s_{W}^{2}\to s_{W}^{2}(1-\Delta)$
\cite{Degrassi:1989ip,Sirlin:2012mh} in \cref{eq:tree}; here $\Delta$
accounts for a subset of 1-loop corrections for eES and
 we are mostly interested in the flavor-dependent contribution $\Delta_{l}\equiv\alpha(6\pi s_{W}^{2})^{-1}\,\text{Log}(m_{W}^{2}/m_{l}^{2})$
where $m_{W}$ and $m_{l}$ are $W$ boson and charged lepton masses, respectively.
$\Delta_{l}$ is evaluated assuming a vanishing momentum transfer
which is a reasonable approximation given the magnitude of the considered
$E_{\nu}$. It turns out that $\Delta_{\mu}-\Delta_{\tau}\approx0.01$,
and this propagates to $\mathcal{O}$(1\%) difference in the cross
section; the $\nu_{\mu}$ cross section is larger than the $\nu_{\tau}$ one,
see the cyan line in \cref{fig:xsec}. In the figure, we also compare
each of these cross sections at 1-loop level with the respective tree-level
expression (see \cref{eq:tree}); these $\mathcal{O}(\alpha)$ effects
are shown by purple and green lines, respectively. Let us stress that
in making \cref{fig:xsec} as well as for our analysis presented
in the next sections, we utilize results from \cite{Tomalak:2019ibg,Hill:2019xqk}.
There, both electroweak and QED corrections as well as the emission
of soft photons is taken into account for eES. Note that, as a cross
check, we also explicitly implemented the expressions from \cite{Sarantakos:1982bp}
and found consistent results; for instance, the difference between
$\nu_{\mu}$ and $\nu_{\tau}$ cross sections was found, for any value of $T$, 
to deviate from cyan line in \cref{fig:xsec} by no more than 0.1\%.

The above discussion on the cross sections is important for the
detection of neutrinos. However, one may also wonder about the impact
of radiative corrections in neutrino propagation; after all, for the
computation of the Mikheyev-Smirnov-Wolfenstein (MSW) matter potential \cite{Wolfenstein:1977ue,Mikheev:1986gs,Mikheev:1986wj}
the same diagrams as those presented in \cref{fig:feynman} should
be evaluated at a zero momentum transfer. The flavor-dependent NLO effects in the propagation were studied in \cite{PhysRevD.35.896}
where it was found that when summing the relevant contributions, including
diagrams with neutrino scattering on both electrons and quarks (nucleons),
there is a cancellation at $\mathcal{O}(\alpha)$ level for neutrinos
traveling through a neutral unpolarized medium. In turn, the flavor-dependent
effects in neutrino propagation arise only at $\mathcal{O}(\alpha(m_{l}^{2}/m_{W}^{2})\text{Log}(m_{W}^{2}/m_{l}^{2}))\approx10^{-6}$
which is rather small. This led the authors of \cite{MINAKATA1999256}
to conclude that such smallness of flavor-dependent terms leads
to virtually no sensitivity to $\delta_{\rm CP}$ when studying solar neutrinos. In this paper we will oppose such a claim and demonstrate the sensitivity to CP violation by utilizing $\mathcal{O}(\alpha)$ differences in eES for different neutrino flavors.

\begin{figure*}[t]
	\centering
	\includegraphics[width=0.95\textwidth]{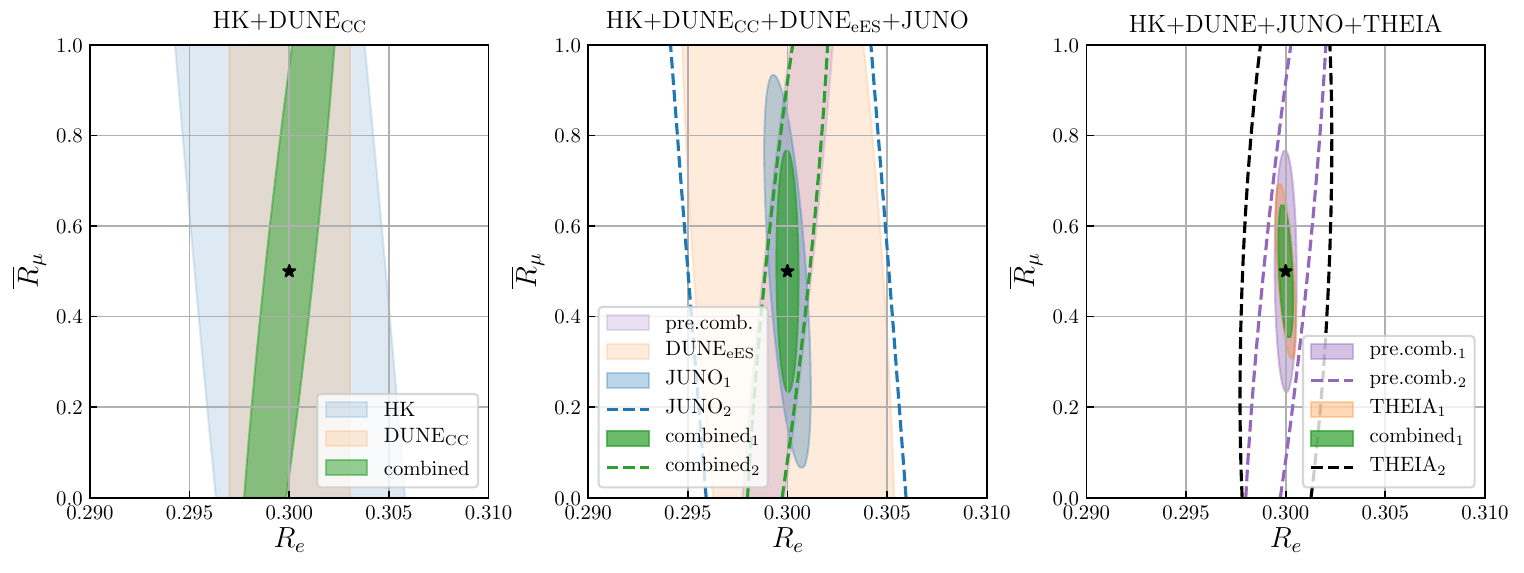}
	\caption{The capability of next-generation neutrino detectors for measuring solar neutrino flavor compositions. The subscripts $1,2$ indicate that JUNO$_{1,2}$   and/or THEIA$_{1,2}$  is included in the analysis---see text for details. All contours are at $1 \sigma$ CL and the assumed true value is marked by $\star$.
	}
	\label{fig:flavor}
\end{figure*}

\textbf{Measuring the Flavor Composition.} Now, let us utilize the flavor-dependent cross sections to scrutinize the potential for measuring the solar neutrino flavor composition $(\phi_{\nu_{e}}:\phi_{\nu_{\mu}}:\phi_{\nu_{\tau}})$
where $\phi_{\nu_{\alpha}}$ denotes the flux of $\nu_{\alpha}$.
For convenience, we define 
\begin{equation}
R_{\alpha}\equiv\frac{\phi_{\nu_{\alpha}}}{\phi_{{\rm total}}}\thinspace,\ \phi_{{\rm total}}\equiv\phi_{\nu_{e}}+\phi_{\nu_{\mu}}+\phi_{\nu_{\tau}}\thinspace.\label{eq:R}
\end{equation}
Since $R_{e}+R_{\mu}+R_{\tau}=1$, only two of the ratios are independent. We also define $\overline{R}_{\mu}\equiv \phi_{\nu_{\mu}}/(\phi_{\nu_{\mu}}+\phi_{\nu_{\tau}})$ which can freely vary between 0 to 1. 
Here, we should stress that the flavor ratios are actually energy-dependent according to the standard MSW solution. Nevertheless, in this section, for demonstration purposes, we take them as energy-independent in the fit and investigate how well the solar neutrino flux components can be measured. This contrasts to the following section in which we study $\delta_{\text{CP}}$ sensitivity in the framework of the standard MSW solution.

To date, only $R_{e}$ has been successfully measured, via combination of eES, neutrino-nucleus charged current (CC) and neutral current (NC) scattering. Among the three channels,  eES and neutrino-nucleus CC scattering have different cross sections for $\nu_{e}$ and $\nu_{x}$ ($x=\mu$ or $\tau$) while neutrino-nucleus NC scattering is flavor independent at the leading order. When combining the three channels, $R_{e}$ is actually overconstrained\footnote{The  data from the three channels turns out to be compatible with each other; see e.g., Fig.~29 of Ref.~\cite{SNO:2005oxr}.} but the flavor composition $\phi_{\nu_\mu}:\phi_{\nu_\tau}$ still cannot be resolved. To measure this flavor composition, one has to include the aforementioned radiative corrections which induce 
small differences between $\nu_{\mu}$ and $\nu_{\tau}$ cross sections.

With the NLO corrections included and by taking $(R_{e},\ \overline{R}_{\mu})$ as free
parameters, we perform a $\chi^{2}$-fit analysis to evaluate
the potential of next-generation neutrino experiments to measure $\phi_{\nu_{e}}:\phi_{\nu_{\mu}}:\phi_{\nu_{\tau}}$.

The next-generation neutrino detectors will feature complementary advantages. The HK detector, which will be a $187$ kt water Cherenkov detector, will have the highest statistics in the eES channel at energies above a certain threshold. We assume that the threshold is the same as for Super-Kamiokande, namely  $T\gtrsim 3.49$ MeV~\cite{Super-Kamiokande:2016yck}. 
The far detector of JUNO will be a $20$ kt liquid scintillator (LS) detector and will also have high statistics in the eES channel. Despite the smaller fiducial mass, its detection threshold, if neglecting cosmogenic backgrounds, could be much lower than the one at HK due to the high light yield in LS. If one only counts the  yield of photoelectrons, the threshold is likely to reach $T \gtrsim 0.1$ MeV\footnote{This is a reasonable assumption, given that KamLAND reached about $0.2$ MeV according to Ref.~\cite{Beacom:2002hs} and JUNO will have a significantly higher yield of photoelectrons with respect to KamLAND.}.
However, cosmogenic backgrounds~\cite{Borexino:2021pyz} pose the main challenge to the detection of low-energy events. In Borexino, these backgrounds are effectively reduced via a few sophisticated methods (e.g. TFC)~\cite{Borexino:2013zhu}, which enable Borexino to successfully detect the low-energy part (pp, $^7$Be, CNO) of solar neutrino spectrum. 
Since the underground depth of JUNO is about half the depth of Borexino, we anticipate that JUNO might be able to apply the same background reduction techniques to some extent. Hence, for JUNO, we consider two cases, denoted by JUNO$_1$ and JUNO$_2$. JUNO$_1$ simply assumes the effective reduction of cosmogenic backgrounds while JUNO$_2$ conservatively assumes that all eES events below 2 MeV are not discernible from the background, leading to a relatively high threshold $T \gtrsim 2$ MeV. 
The far detector of the DUNE experiment will contain $40$ kt of liquid Ar and it is anticipated to measure solar neutrinos in both CC ($\nu_{e}+{}^{40}\text{Ar}\to e^{-}+{}^{40}\text{K}$, denoted by ArCC)
and eES channels~\cite{Capozzi:2018dat}. We assume that the threshold of ArCC process at DUNE will be $E_\nu \gtrsim 7$ MeV. Very recently, the THEIA
experiment has been proposed~\cite{Theia:2019non}, with a $100$ kt water-based liquid scintillator (WbLS) detector placed deep underground. A high (low) percentage of LS in WbLS would decrease (increase) its capability of measuring the direction of interacting neutrinos, but it would lead to a lower (higher) energy threshold. 
If pure LS is employed, it would be very similar to Borexino, which according to Refs.~\cite{Borexino-talk,Bellini:2016kau} has detected pp neutrinos successfully with the threshold around 150-200 keV.  If 5\% WbLS is used, then it is likely that the threshold may reach 0.6 MeV~\cite{Theia:2019non}.
As the percentage of LS is not determined yet, we consider two configurations for THEIA, namely THEIA$_1$ using pure LS with a 0.1 MeV threshold, and THEIA$_2$ using 5\% WbLS with a 0.6 MeV threshold. 
Dark matter detectors could contribute in the NC channel by collecting coherent elactic neutrino-nucleus scattering (CE$\nu$NS) events induced by solar neutrinos. However, the statistics of such NC events in ton-scale detectors is low (in particular, they have not been measured to date), compared to the achieved SNO (kt-scale) observations of $\nu+{}^{2}\text{H}\to\nu+n+p$ which
is also a NC channel. The contribution of dark matter detectors for distinguishing between solar fluxes of $\phi_\mu$ and $\phi_\tau$ is therefore expected not to be competitive to the above introduced experiments and channels.  
\begin{figure*}[t!]
	\centering
	\includegraphics[width=0.99\textwidth]{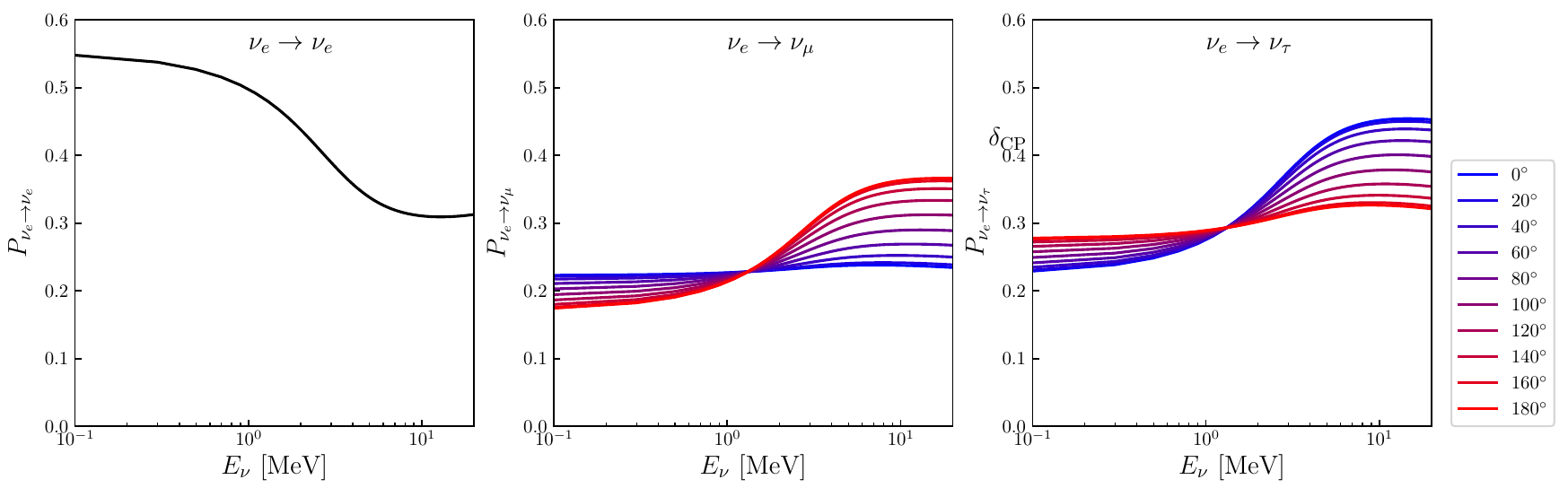}
	\caption{The solar neutrino transition probabilities $P_{\nu_e\to \nu_{\alpha}} (E_{\nu})$ for various values of $\delta_{\rm CP}$. 
		\label{fig:P}}
\end{figure*}
In our analysis,  the eES channel at HK, DUNE, JUNO and THEIA is studied and combined with the ArCC channel at DUNE. 

The eES event rate reads\footnote{In this work we are only concerned with the electron energy spectrum and we do not consider the angular spectrum which can be measured at HK and THEIA. Including the angular spectrum could in principle further improve the results.}
\begin{equation}
\frac{dN}{dT}= N_{e}\Delta t\sum_{\alpha}\int\frac{d\sigma_{\nu_{\alpha}}}{dT}\left(T,E_{\nu}\right)\phi_{\nu_{\alpha}}(E_{\nu})dE_{\nu}\thinspace,\label{eq:dNdT}
\end{equation}
where $N_{e}$ denotes the number of electrons in
a detector and $\Delta t$ is the exposure time and we assume $\Delta t=10$ years for all the considered experiments. For ArCC, in which $T$ is also the main observable, the event rate is computed in a similar way, except that $N_{e}$ should be replaced by the number of argon nuclei. In our analysis, we also assume that the uncertainties on solar neutrino flux components \cite{Bergstrom:2016cbh} will improve and reach $\sim 1$\%\footnote{We also implicitly assume that systematic uncertainties related to the detector are subleading. For instance, the present systematical uncertainties for Super-Kamiokande detector are already at $\mathcal{O}(1\%)$ level \cite{Super-Kamiokande:2016yck}, and we anticipate further improvements for HK.}. The treatment of such uncertainties is incorporated by $\phi_{\rm total}\to (1+a) \phi_{\rm total}$ where the normalization factor, $a$, is included with an uncertainty of $\sigma_a=1\%$ and marginalized over in the $\chi^2$ analysis.

For each channel and each experiment, we perform a binned $\chi^2$ analysis to assess the sensitivity to the flux ratios  $(R_{e},\ \overline{R}_{\mu})$ assuming the true value is $(0.3,\ 0.5)$. 
In Fig.~\ref{fig:flavor}, we show the results by considering several experiments individually as well as their combinations. The left panel shows how well the fluxes can be measured by using eES in HK (light blue) and ArCC in DUNE (light brown) as well as their combination (green). In the other two panels we add more experiments; the ``pre.comb.'' label in the legend of a given panel refers to the combination (green region) from the panel to the left. If all next-generation experiments are combined, 
and if the LS experiments JUNO and THEIA can reach the optimal configurations of $T=0.1$ MeV (denoted by subscript 1 in Fig.~\ref{fig:flavor}) we expect that $\overline{R}_{\mu}$ could be measured to the precision of $0.5\pm 0.15$. In such a case, the majority of neutrino events  would arise from the interactions of solar pp neutrinos with $E_\nu \lesssim 0.4$ MeV.

\begin{figure}[t]
\centering
\includegraphics[width=8cm]{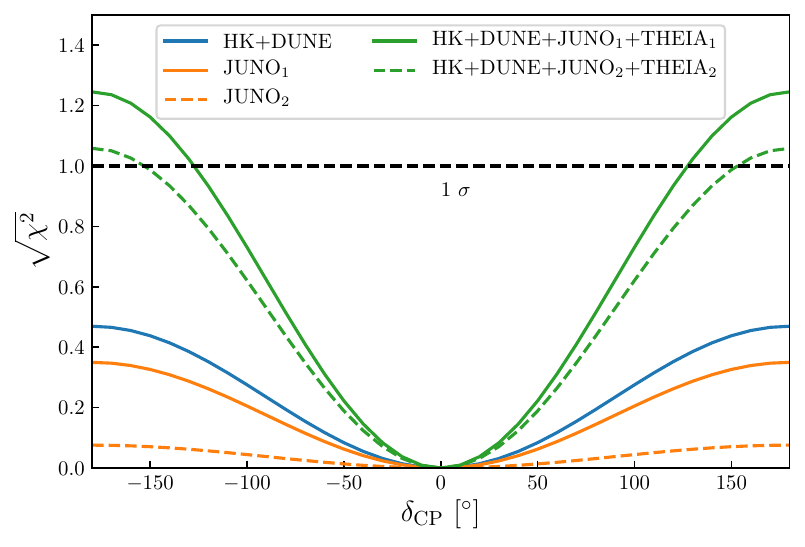}
\caption{$\delta_{\rm CP}$ sensitivity in solar neutrino measurements at next-generation neutrino experiments.}
\label{fig:sens}
\end{figure}

\textbf{Solar neutrinos as a probe of CP violation.}
In Fig.~\ref{fig:P}, we show $P_{\nu_e\to \nu_{\alpha}}(E_{\nu})$  for various $\delta_{\rm CP}$ values, obtained using the adiabatic approximation (for the range of its validity, see detailed discussion in \cite{Xu:2022wcq}). As shown in the figure,  $P_{\nu_e\to \nu_{\mu}}$ and  $P_{\nu_e\to \nu_{\tau}}$  vary significantly for $\delta_{\rm CP} \in [0,\pi]$; the variation can be as large as $\sim 50 \%$. 
In making the figure, we have also included the matter effect for neutrinos propagating inside Earth; specifically, we have averaged over the day and the night values of $P_{\nu_e\to \nu_{\alpha}}$. 
Given the previous conclusion that the full flavor composition of solar neutrinos can be measured using NLO cross sections (see again \cref{fig:flavor}), we expect that precision measurements of solar neutrinos should exhibit sensitivity to $\delta_{\rm CP}$. To demonstrate that, we performed a $\chi^2$ analysis and the result is presented in Fig.~\ref{fig:sens}. As shown in the figure, the full combination of next-generation detectors allows  $\delta_\text{CP}=0$ and $\delta_\text{CP}=\pi$ ($\pi/2$) to be differentiated at $\sim 1\sigma$ ($\sim 0.5\sigma$) CL. We also found that, for $\sigma_a \lesssim 0.1$\%, $2\sigma$ can be reached.

In making Fig.~\ref{fig:sens}, we took $\delta_{\rm CP}$ as the only fitting parameter and we fixed all other oscillation parameters at their present best fit values \cite{Esteban:2020cvm}. The reduction of uncertainties on the mixing angles, especially $\theta_{23}$, is anticipated across relatively short time scales \cite{Song:2020nfh}. Currently, varying $\theta_{13}$, $\theta_{12}$, and $\theta_{23}$ within their $1\sigma$ ranges \cite{Esteban:2020cvm} would cause $P_{e\mu}$ to change by $0.2\%$, $2\%$, $4\%$, respectively. We leave a dedicated investigation on the correlation between the mixing angles and $\delta_{\rm CP}$ to future work.

\textbf{Summary and Conclusions.} 
The survival probability of solar $\nu_e$ has been measured at a number of experiments in the $\sim 1$--$10$ MeV energy range. On the other hand, directly measuring solar $\nu_\mu$ and $\nu_\tau$ fluxes is much more challenging and has not been performed to date. In this work, we propose a viable method to measure the flavor composition of solar neutrinos by utilizing differences between $\nu_\mu$ and $\nu_\tau$ eES cross sections. While these cross sections are identical at tree level, radiative corrections involving $\mu$ and $\tau$ leptons in loop diagrams induce an $\mathcal{O}(1\%)$ difference. This allows one to probe the flavor composition of solar neutrinos through the observation of electron recoil spectrum. We have quantified such an effect by studying the potential of several forthcoming neutrino detectors including HK, DUNE and JUNO, together with the proposed THEIA experiment.  If THEIA and JUNO can realize 0.1 MeV detection threshold, then the ${\cal O}(1\%)$ difference in the cross sections would allow the combination of these experiments to effectively resolve the full flavor composition, see Fig.~\ref{fig:flavor}.

Furthermore, since $P_{\nu_e\to \nu_{\mu}}$ and $P_{\nu_e\to \nu_{\tau}}$ depend on $\delta_{\rm CP}$, the leptonic CP violation could be probed with solar neutrinos if the flavor-dependent radiative corrections are taken into consideration. We have assessed the sensitivity to $\delta_{\rm CP}$ and found that $\sim 1 \sigma$ CL can be reached in differentiating between $\delta_{\rm CP}=0$ and $\delta_{\rm CP}=\pi$. 
This result would improve to $\sim 2\sigma$ provided that the uncertainties in solar neutrino flux reach sub-percent level.  
For such a measurement, the main experimental challenge would be low-threshold detection of solar neutrinos. Hence next-generation LS experiments like JUNO and THEIA will be particularly important.

\textbf{Note Added.} 
As we were finalizing this work, Ref. \cite{Mishra:2023jlq} appeared on arXiv. There, the authors scrutinize the prospects for measuring NLO effects with CE$\nu$NS at next-generation dark matter experiments. While both CE$\nu$NS and neutrino-electron scattering feature very small uncertainties in the cross section, the latter process has already been measured with large statistics while the coherent elastic neutrino-nucleus scattering is yet to be recorded for solar neutrinos. Hence, we regard neutrino-electron scattering channel as more promising for measuring the flavor composition of solar neutrinos at next-generation experiments.

\textbf{Acknowledgments.} 
We would like to thank Leonardo Ferreira and Oleksandr Tomalak for
very useful discussions. X.J.X is supported in part by the National
Natural Science Foundation of China under grant No. 12141501 and also supported by CAS Project for Young Scientists in Basic Research 
(YSBR-099). 
X.J.X would also like to thank CERN for the hospitality and the financial support 
during his visit when this work was performed in part.

\bibliographystyle{JHEP}
\bibliography{refs}

\end{document}